# Understanding Visually Impaired Tramway Passengers' Interaction with Public Transport Systems

**Dominik Mimra[1], Dominik Kaar[1], Enrico Del Re[2], Novel Certad[2], Joshua Cherian Varughese[2], David Seibt[1] and Cristina Olaverri-Monreal[2]**

[1] Institute for Sociology, Johannes Kepler University Linz, Austria

[2] Department Intelligent Transport Systems, Johannes Kepler University Linz, Austria

**ABSTRACT**

Designing inclusive public transport services is crucial to developing modern, barrier-free smart city infrastructure. This research contributes to the design of inclusive public transport by considering accessibility challenges emerging from socio-technical systems, thus demanding the integration of technological and social solutions. Using Actor-Network Theory (ANT) as a theoretical framework and a mixed-method approach, including shadowing and a focus group, this study examines the socio-technical networks that shape accessibility experiences for visually impaired passengers utilizing the tram in Linz, Austria. Key dimensions that influence public transport accessibility are identified: network configuration, mobility patterns, technology integration, and warning systems. The results show that accessibility emerges from complex interactions between human actors (passengers, staff) and non-human actors (assistive devices, infrastructure) rather than being an inherent property of transport systems. Digital technologies serve multiple functions, from navigational assistance to broader social inclusion, although users' comfort with technology varies. Participants emphasized the importance of the "two-sense principle" for warning signals, with directional audio and tactile feedback particularly valuable.

# INTRODUCTION

Disabled people depend more on public transportation than the general population (Penfold et al., 2008). Therefore, their quality of life is affected by the availability of adequate public transport (Bezyak et al., 2017). The absence or inaccessibility of public transport could result in socio-economic isolation and marginalization of disabled people (Chenoweth & Stehlik, 2004). However, despite the importance of mobility, most persons with disabilities experience difficulties accessing public transportation (Bezyak et al., 2020).

Visually impaired individuals face significant challenges when navigating public transportation systems, which impacts their independence and mobility. Most notably, access to information (e.g., identifying the correct bus, tram, etc.), the drivers' unawareness of the needs of visually impaired people, vehicle design, and technological barriers such as unreliable voice announcements or unreliable warning signals affect their ability to use public transport (Boadi-Kusi et al., 2023). Addressing these issues is essential for creating inclusive, equitable public transportation systems for individuals with visual impairments.

We use actor-network theory (ANT) as our primary framework for exploring the accessibility of public transport for visually impaired people. ANT (Latour, 2005) examines the distribution of agency between human and non-human actants in socio-technical networks. A social-technical network is a dynamic system of heterogeneous elements—including humans, technologies, organizations, policies, and physical infrastructure—dynamically linked to each other through translation processes. Since agency in these networks is distributed rather than concentrated in individual actors, both ability and disability are viewed as resulting from the dynamic relations between human and non-human entities (Moser, 2006). Using ANT thus allows us to consider accessibility as an emergent property of networked interactions rather than a fixed characteristic of infrastructure (Galis & Lee, 2014; Velho, 2021). To explore these interactions empirically, we collaborated with the "Hilfsgemeinschaft—Austrian Association for the Blind and Visually Impaired." We investigated barriers such as inaccessible information and increased safety risks. The findings emphasize the importance of considering the diversity of user groups for developing inclusive public transport systems and policies.

Two fundamental concepts of ANT drive our analysis: actants and translation. Actants are all entities that make a difference in the effects produced by a socio-technical network (Latour, 2005:71). The concept allows us to analyze the distribution of agency in the network and to symmetrically describe the influence of human (e.g., transport staff, companions) and non-human actants (e.g., white canes, smartphones) on accessibility outcomes. Translation describes the processes of aligning actants into networks (Latour, 2005:106f). The term highlights that, like linguistic translation between words, forging associations between heterogeneous actants involves subtle transformations and often requires mediation by additional actants. Such mediation processes are crucial for accessibility, e.g., when tactile paving converts spatial information into tactile signals or when smartphone apps translate text into audible formats. Some actants, such as ticket validation systems or tram platforms, take on special significance for network coherence, as they form "obligatory passage points" (Callon, 1984) with which other actants must align to produce the effect of accessibility. Such obligatory passage points are thus critical for the overall accessibility or inaccessibility of public transport systems.

Building on the view of public transport systems as socio-technical networks, our framework further includes the concepts of script and mental model, which specify the design and use of technical objects and infrastructures as key elements of the networks we study. The notion of script explains how technologies may come to embody patterns of inclusion and exclusion. Akrich (1992) argues that designers inscribe assumptions about future users into the material form of the technologies they develop. As a result, technologies like trams or smartphone apps often privilege certain user groups while disadvantaging others (e.g., visually impaired people). However, scripts remain open to interpretation, and users may find creative ways to overcome barriers. Visually impaired individuals, for instance, develop mental models (Norman 1988:31f) of the technologies and infrastructures they regularly use, enabling them to develop effective strategies for navigating public transport systems. Accordingly, the accessibility of public transport is not a fixed property but is shaped by users' capacity to learn and adapt within socio-technical systems.

In the remainder of this paper, we use this framework to examine a public transport system in Linz, Austria. We identify where barriers emerge and develop strategies for improving accessibility for visually impaired users. To this end, the paper is structured as follows: Section 2 details our methodology, including participant recruitment, ethical considerations, focus group design, and qualitative data analysis techniques. Section 3 presents the results, examining network configuration, mobility patterns, technology integration, warning systems, and additional dimensions influencing visually impaired passengers' experiences. Finally, Section 4 concludes the paper with key insights, implications for public transport design and policy, and suggestions for future research.

## METHODOLOGY

Building on the theoretical framework outlined above, our methodology seeks to capture how the configuration of socio-technical networks shapes the accessibility of public transport systems for visually impaired passengers. To this end, a shadowing observation and a focus group session were conducted in November 2024 to uncover how passengers perceive relationships with various human and non-human actors in the network.

### Participant Recruitment and Ethical Considerations

The participants were recruited with the help of the "Austrian Association for the Blind and Visually Impaired" (In German: Hilfsgemeinschaft der Blinden und Sehschwachen Österreichs). The Association's involvement ensured that the research methods appropriately addressed the needs of visually impaired individuals. The selection of participants represented diverse degrees of visual impairment and varying levels of public transport experience. The participants also exhibited different levels of familiarity with assistive technologies and varying degrees of independence during travel, from fully independent travel to needing an accompanying person for the entire route.

We ensured that ethical considerations related to the visual impairment of participants were addressed throughout the research process. All participants provided informed consent using consent forms available in standard and accessible formats. Guillemin and Gillam's (2004) framework for ethical research practice was followed, and the Association's office manager reviewed the research protocol.

The final sample comprised seven visually impaired individuals and the companion of one person. The age range was between 22 and 74 years. Five of the participants were male, and three were female. The degree of visual impairment varied from severe visual impairment to complete blindness from birth.

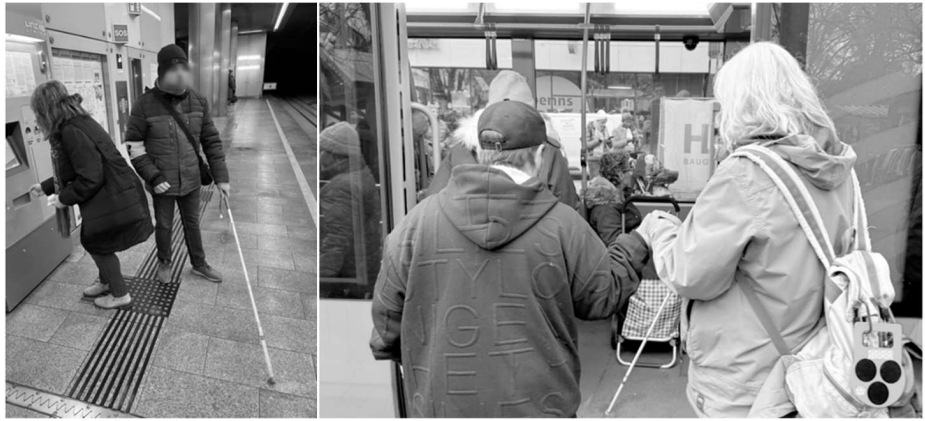

**Figure 1:** Visually impaired participant with a companion when buying the ticket (left). A passer-by helps a participant to board a tram (right).

### Shadowing

Our methodology also included shadowing observations, as described by Gill et al. (2014). It involved shadowing visually impaired participants as they navigated the tram system, documenting their interactions with human and non-human actants without direct intervention. The shadowing protocol captured participants' spatial navigation techniques, interactions with the physical infrastructure, use of technological aids, communication strategies, and responses to unexpected situations. This observational data complemented the focus group findings, allowing us to witness real-time navigation strategies and socio-technical interactions in their natural context.

### Focus Group

We structured the session following Krueger and Casey's (2015) guidelines for inclusive discussion environments. The session lasted one hour and fifteen minutes and was moderated by a sociologist with previous experience conducting focus groups. The discussion progressed from general experiences with public transport to specific discussions of technological solutions and social interactions. The moderator used techniques to manage group dynamics in discussions involving visually impaired participants, ensuring equal participation and clear communication. The session was recorded (audio and video) with the participants' consent, and additional note-takers documented nonverbal interactions and group dynamics.

### Qualitative Data Analysis

The collected data and supplementary documentation of group dynamics were processed and interpreted using Miles, Huberman, and Saldana's (2014) framework for qualitative data analysis. The analysis proceeded through several phases, beginning with the transcription of audio recordings and digitization of field notes. All materials were imported into MaxQDA software for systematic coding and analysis. Coding was done using Mayring's (Mayring & Fenzl, 2019) qualitative content analysis methodology, incorporating deductive and inductive coding approaches. We derived initial coding categories from ANT concepts and added additional categories through iterative data analysis. We resolved interpretive discrepancies during the analysis and refined the analytical framework (Table 1 depicts the code system). This approach ensured the reliability of the findings through investigator triangulation.

**Table 1**: Final code system used for the qualitative analysis.

| 1. Mobility patterns | 2. Dependence on Public Transport | 3. Role of Technology |
|---|---|---|
| 1.1 Frequency of Use | 2.1 Absolute Dependence | 3.1 Assistive Devices |
| 1.2 Purpose of Use | 2.2 Conditional Dependence | 3.2 Smartphone Applications |
| 1.3 Alternative Means of Transport | | 3.3 Communication and Social Media |
| | | 3.4 Technology Attitudes |
| **4. Barriers and Limitations** | **5. Self-Reliance and Independence** | **6. Inclusion** |
| 4.1 Physical Barriers | 5.1 Enhanced Independence | 6.1 Interaction with Society |
| 4.2 Technical Failures | 5.2 Dependence on Technology | 6.2 Overcoming Isolation |
| 4.3 Organizational Issues | 5.3 Learning Strategies | 6.3 Risk Management |
| 4.4 Spatial Navigation | | |
| **7. Suggestions for Improvement** | **8. Comparisons with Other Systems** | **9. ANT analysis** |
| 7.1 Technological Enhancements | 8.1 Regional Comparisons | 9.1 Distributed Action and Networks |
| 7.2 Infrastructure Adjustments | 8.2 International Comparisons | 9.2 Actants (Non-human Actors) |
| 7.3 Policy Changes | | 9.3 Communication Strategies |
| | | 9.4 Delegation |
| | | 9.5 Script and Inscription |
| | | 9.6 Programs and Anti-Programs |

## ANALYSIS RESULTS

Our findings comprise results on multiple dimensions. Table 2 lists selected empirical evidence. Section 3.5 adds further dimensions.

### Network Configuration and Dynamics

Interaction of human (visually impaired passengers, transport staff, and fellow travelers) with technical actants (assistive devices, infrastructure) resulted in accessibility outcomes. One participant noted: "For me, technology plays a very big role. Especially the transmitter for the trams and traffic lights is really great" (S05, pos. 33).

Further findings from the focus group session showed that translation processes between these actants emerged through various mechanisms. Smartphone apps translate visual information into audio output, while staff mediate between institutional systems and individual needs. These processes demonstrate that agency is distributed in socio-technical networks (Latour 2005), and accessibility emerges from assemblages of human and non-human entities rather than being an inherent property of transportation infrastructures.

### Mobility Patterns and Transportation Usage

Regular users demonstrated sophisticated knowledge of the transport system. The data shows strong correlations between frequency of use and enhanced independence, with some participants deliberately using public transport to develop their navigational skills: "I use it consciously also towards Linz station, just to train" (S05, Line 2). Norman's (1988) concept of mental models applies here, as visually impaired passengers enhance navigation skills through repeated engagement with transportation systems. Most participants had limited alternative transport options, emphasizing the critical importance of accessible public transport systems.

**Technology Integration and Digital Solutions**

The study showed that digital technologies serve multiple functions, from practical navigation assistance to broader social inclusion. For example, the "Wann Today" app instantly provides real-time public transportation departure times for buses, trams, and trains in several cities through an accessible widget featuring voice assistance technology that can read out all screen content for visually impaired users through gesture controls. One participant commented: "The apps on the smartphone I find quite important. Departure information of public transport, like the train app or for the transport association" (S03, pos. 37). The findings were in line with strong correlations emerging between technology use and enhanced independence, though comfort levels varied. Callon's (1984) translation processes are evident when smartphone apps convert real-time tram data into audio cues, aiding independence while revealing varying user comfort with technology.

**Warning Signals**

Participants described ideal warning solutions based on the "two senses principle" combining acoustic and tactile feedback. Hearing was crucial, with the signal origin being critical: "I am in favor of the signal coming from where the danger is coming from [...]" (S05, pos. 143).

The concept of a "smart cane" was mentioned as a potential future technology: "My consideration would be [...] that perhaps the chip is built into the long cane, which then emits a signal or a vibration" (S06, pos. 132).

**Summary of Findings and Additional Dimensions**

In addition, our analysis identified the following relevant dimensions that influence the experience of visually impaired passengers:

- **Accessibility Barriers and Enablers:** Physical, social, and technical barriers create overlapping challenges

- **User Experience and Adaptation Strategies:** Sophisticated combinations of technical aids and social strategies maintain independence

- **Infrastructure and System Design:** Physical design significantly impacts accessibility outcomes

- **Policy and Organizational Factors:** Institutional policies shape accessibility through both intentional supports and unintended barriers

- **Risk Management and Safety:** Users develop complex strategies balancing safety with independence

- **Spatial Navigation:** Users rely on memorized patterns and environmental cues

These dimensions indicate that accessibility emerges from the alignment of diverse socio-technical network elements and requires integrated approaches to improving it.

**Table 2**. Selected Empirical Evidence

| Dimension | Empirical Evidence | Implication |
| --- | --- | --- |
| Technology Integration | "Without the apps, my life would be much more difficult" (S03, pos. 43) | Digital solutions need careful implementation, considering varying user comfort levels. |
| Social Interaction | "I simply address people: 'Please let a blind person sit here'" (S03, pos. 187) | Human assistance continues alongside technical solutions. |
| Infrastructure Design | "If the tram is packed, you cannot reach your priority space" (S03, pos. 205) | Physical design requires consideration of diverse user needs. |
| Regular System Use | "I use it consciously also towards Linz station, just to train" (S05, pos. 2) | Familiarity builds confidence and independence. |
| Warning Signals | "The signal [should come] from where the danger is coming from" (S05, pos. 143) | Directional signals are crucial for hazard assessment. |

## CONCLUSIONS AND FUTURE WORK

This research has investigated how visually impaired passengers interact with urban tram systems, using the Actor-Network Theory to understand accessibility as an emergent property of socio-technical networks. Our findings reveal key insights with important implications for public transport design and policy.

First, our research shows that accessibility is not an inherent property of transport systems but emerges from the complex relations between human (visually impaired passengers, transport staff) and non-human actants (assistive devices, infrastructure). This finding is consistent with the distributed agency perspective of Actor-Network Theory, which helps reframe accessibility challenges from individual limitations to network effects. By applying this framework, we identified critical interaction points—such as platform-vehicle transitions—where multiple actors must successfully converge to ensure accessibility.

Second, our findings show that digital solutions provide essential but varied support for visually impaired passengers. Applications such as 'Wann Today' transform visual information into accessible audio formats, significantly increasing independence. However, their effectiveness varies considerably depending on the technological comfort and expertise of the user. This variability underscores the importance of designing solutions to accommodate users with diverse bodily characteristics and levels of technological proficiency.

Third, physical design and infrastructure remain essential factors for accessibility outcomes. Critical nodes, such as platform-vehicle transitions, function as 'obligatory passage points' where systemic vulnerabilities are exposed. These transitions reveal where multiple entities in the network must successfully converge for accessibility.

Fourth, warning systems designed according to the "two senses principle" are particularly valuable. Participants emphasized that directional signals indicating the origin of potential hazards significantly improve safety, allowing users to orient themselves better relative to approaching hazards.

Finally, our research underlines the continuing importance of social interactions alongside technological solutions. Human assistance remains essential to digital and infrastructural improvements, with one participant relying on a human companion and many others developing sophisticated strategies for engaging fellow travelers when needed.

These findings have important implications for the OptiPex project and broader public transport design. They suggest that effective solutions must design and implement technological innovations with careful consideration of diverse user needs and existing adaptation strategies. By applying ANT's focus on network adaptation, transport providers can identify critical points of intervention where modest changes could lead to significant improvements in accessibility.

Future research could explore how emerging technologies, such as AI and the Internet of Things, could further improve the accessibility of public transport while remaining mindful of potential exclusion. Longitudinal studies examining how visually impaired passengers develop and refine navigation strategies over time would also provide valuable insights into the learning dimensions of accessibility.

In conclusion, this research contributes to more inclusive public transport systems by highlighting how accessibility emerges from the alignment of different socio-technical elements. By considering the complex interplay between human actors, technological solutions, and physical infrastructure, transport providers can develop more comprehensive approaches to accessibility that benefit all users.

## ACKNOWLEDGEMENTS

This work has been conducted as a part of the OptiPEx project (No.101146513) funded by the European Union. However, the views and opinions expressed are those of the author only and do not necessarily reflect those of the European Union or CINEA. Neither the European Union nor the granting authority can be held responsible.